\documentclass[journal]{IEEEtran}
\usepackage{filecontents}
\usepackage{caption}
\usepackage[noadjust]{cite}
\usepackage{amsthm}

\ifCLASSINFOpdf
\usepackage[pdftex]{graphicx}
\else
\fi
\usepackage{amsmath}
\usepackage{amssymb}
\usepackage{mathtools}
\DeclarePairedDelimiter\floor{\lfloor}{\rfloor}
\newcommand{\RNum}[1]{\uppercase\expandafter{\romannumeral #1\relax}}
\usepackage{adjustbox}
\newtheorem{theorem}{Theorem}

\usepackage{subcaption}
\usepackage{mwe}
\usepackage{algorithmic}

\usepackage{array}
\usepackage{makecell}
\usepackage{multirow}
\newcolumntype{M}[1]{>{\centering\arraybackslash}m{#1}}
\newcolumntype{P}[1]{>{\centering\arraybackslash}p{#1}}

\begin{document}
%
\title{On Complex Conjugate Pair Sums and Complex Conjugate Subspaces}
%
%
%

\author{Shaik~Basheeruddin~Shah,~\IEEEmembership{Student~Member,~IEEE,}
~Vijay~Kumar~Chakka,~\IEEEmembership{Senior~Member,~IEEE,}~and~Arikatla~Satyanarayana~Reddy 
\thanks{Shaik Basheeruddin Shah and Vijay Kumar Chakka are with the Department of Electrical Engineering, Shiv Nadar University, India (e-mail: bs600@snu.edu.in, Vijay.Chakka@snu.edu.in).}
\thanks{Arikatla Satyanarayana Reddy is with the Department of Mathematics, Shiv Nadar University, India (e-mail: satyanarayana.reddy@snu.edu.in).}}
\maketitle

\begin{abstract} In this letter, we study a few properties of Complex Conjugate Pair Sums (CCPSs) and Complex Conjugate Subspaces (CCSs).
Initially, we consider an LTI system whose impulse response is one period data of CCPS. For a given input $x(n)$, we prove that the output of this system is equivalent to computing the first order derivative of $x(n)$.
Further, with some constraints on the impulse response, the system output is also equivalent to the second order derivative. 
With this, we show that a fine edge detection in an image can be achieved using CCPSs as impulse response over Ramanujan Sums (RSs). 
Later computation of projection for CCS is studied. Here the projection matrix has a circulant structure, which makes the computation of projections easier. Finally, we prove that CCS is shift-invariant and closed under the operation of circular cross-correlation.
\end{abstract}

%
\IEEEpeerreviewmaketitle
\section{Introduction}
\IEEEPARstart{F}{or} a given $q{\in}\mathbb{N}$, define a set of complex exponential sequences as $H_q = \{S_{q,k}(n) = e^{\frac{j2{\pi}kn}{q}}|0{\leq}k{\leq}q-1, (k,q)=1\}$, where $0{\leq}n{\leq}q-1$ and $(k,q)$ denotes the greatest common divisor (gcd) between $k$ and $q$. 
By adding all the elements of $H_q$, mathematician Srinivasa Ramanujan introduced a trigonometric summation called as Ramanujan Sum (RS) in $1918$, denoted as $c_q(n)$ \cite{Ramanujan}.
Later in 2014, P. P. Vaidyanathan introduced a finite length signal representation using $c_q(n)$ and its circular shifts \cite{6839014}, \cite{6839030}.


Motivated by this, we introduced a trigonometric summation known as Complex Conjugate Pair Sum (CCPS) in one of our previous works \cite{Shah}. 
As $(k,q)$ $=$ $(q-k,q)$, for every $S_{q,k}(n){\in}H_q$, there exists a complex conjugate sequence $S_{q,q-k}(n){\in}H_q$, both together form a complex conjugate pair. 
CCPS $(c_{q,k}(n))$ is defined by adding each complex conjugate pair, i.e.,
\begin{equation}
c_{q,k}(n) = 2Mcos\left(\frac{2{\pi}kn}{q}\right),
\end{equation}
where \footnotesize$M=\begin{cases}
\frac{1}{2}&\ \text{if } q=1\text{ (or) }2\\
1&\ \text{if } q{\geq}3 
\end{cases}$\normalsize, $k{\in}{\hat{U}_q}$ if $q>1$ and $k=1$ if $q = 1$, refer \textit{Notations} for ${\hat{U}_q}$. 
Recently, S. W. Deng et al., introduced a two-dimensional subspace spanned by a complex conjugate pair $\{S_{q,k}(n),S_{q,q-k}(n)\}$ known as Complex Conjugate Subspace (CCS) \cite{7544641}, denoted as $v_{q,k}$. 
In \cite{Shah}, we provided a new basis for CCS using CCPS.
Further, a finite length signal is represented as a linear combination of signals which belong to CCSs known as Complex Conjugate Periodic Transform (CCPT) \cite{Shah}.

Inspired from \cite{6839014,6839030,7964706}, in this letter we discuss several properties of CCPSs and CCSs, which may find  applications in signal processing.
Contributions of this letter can be divided into two parts. 
In the first part, we show that
the operation of linear convolution between a given signal $x(n)$ and $\bar{c}_{q,k}(n)$ is equivalent to computing the first derivative of $x(n)$, where $\bar{c}_{q,k}(n)$ denotes one period data of ${c}_{q,k}(n)$. 
This operation is also equivalent to the second derivative if we consider an odd number $q$ and a circular shift of $\frac{q-1}{2}$ for $\bar{c}_{q,k}(n)$. 
Then, the problem of edge detection in an image is addressed using this derivative equivalent operation. 
Moreover, we compare these results with the results obtained by using RSs.

In the second part, we prove the following properties:
\begin{itemize}
\item CCS is a shift-invariant subspace.
\item Since CCPT is a non-orthogonal transform \cite{Shah}, we
compute the projections onto CCSs. 
Further, we show that the projection matrix is a circulant matrix, which reduces the computational complexity of projections.
\item CCS is closed under circular cross-correlation. Moreover, the circular auto-correlation of any finite length sequence is equal to the weighted linear combination of its projections auto-correlation.
\end{itemize}

The structure of this letter is as follows: Using CCPSs as derivatives and the problem of detecting edges in an image are studied in Section \RNum{2}.
Properties of CCSs are discussed in Section \RNum{3}. Finally, conclusions are drawn in Section \RNum{4}.
 
\textit{Notations:}
$(a,b)$ indicates the gcd between $a$ and $b$. A least common multiple is denoted as $lcm$.
Rounding the value $a$ to the greatest integer less than or equal to $a$ is denoted as $\floor*{a}$.
For a given $n\in\mathbb{N}$, Euler's totient function $\varphi(n)$ is defined as $\varphi(n) = \sum\limits_{i=1}^{n}\floor*{\frac{1}{(i,n)}}$. As $(k,n) = (n-k,n)$, $\varphi(n)$ is even for $n\geq 3$. 
Symbol $d|N$ denotes that $d$ is a divisor of $N$. 
Define a set $\hat{U}_n = \{a{\in}\mathbb{N}\ |\ 1{\leq}a{\leq}\floor*{\frac{n}{2}}, (a,n)=1\}$, hence $\#\hat{U}_n = \frac{\varphi(n)}{2}$.
$M_{m,n}(\mathbb{C})$ indicates set of all $m\times n$ matrices with entries from complex numbers. If $m=n$, $M_{m,n}(\mathbb{C})=M_n(\mathbb{C}).$
\section{CCPSs as Derivatives and Application}
Here we perform an operation using CCPSs, which is equivalent to the derivative. In particular, we prove the following:
\begin{theorem}
Consider an LTI system whose impulse response is $\bar{c}_{q,k}(n),\ q>1$, then for a given input $x(n)$  the output $y(n)$ of the system, i.e., $x(n)*\bar{c}_{q,k}(n)$ is equivalent to computing the first order derivative of $x(n)$.
\label{Th1}
\end{theorem}
\textit{Proof:} Let $x(n) = C$, where $C$ is a constant value, then
\begin{equation}
y(n) = C\sum\limits_{l=0}^{q-1}\bar{c}_{q,k}(l) = 0,\ \text{as }\sum\limits_{l=0}^{q-1}\bar{c}_{q,k}(l) = 0,\ \text{for }q>1.
\end{equation}
If $x(n) = u(n-n_0)$, where $u(n)$ is an unit step sequence, then 
\begin{equation}
y(n) = \sum\limits_{l=0}^{q-1}\bar{c}_{q,k}(l)u(n-n_0-l),
\end{equation}
here $y(n){\neq}0,\ \forall\ n_0{\leq}n{\leq}{n_0+q-2}$. If $x(n) = n$, then
\begin{equation}
\footnotesize
\begin{aligned}
y(n) &= n\underbrace{\sum\limits_{l=0}^{q-1}\bar{c}_{q,k}(l)}_{=0}-\underbrace{\sum\limits_{l=0}^{q-1}l\bar{c}_{q,k}(l)}_{\mathbf{P}}\\
&= -M\left[\sum\limits_{l=0}^{q-1}le^{\frac{j2{\pi}kl}{q}}+\sum\limits_{l=0}^{q-1}l e^{\frac{-j2{\pi}kl}{q}}\right]\\
&= M\left[\frac{q}{1-e^{\frac{j2{\pi}k}{q}}}+\frac{q}{1-e^{\frac{-j2{\pi}k}{q}}}\right] = Mq.
\end{aligned}
\label{Der1}
\end{equation}
\normalsize
From the above analysis, we draw the following conclusions regarding the system output. That is, the system output is:
\begin{enumerate}
\item Zero for constant input.
\item Non-zero at the on transient of the unit step sequence.
\item Non-zero constant along the ramps.
\end{enumerate}
In the \textit{context of image processing}, any function/operation satisfying the above three properties is equivalent to first order derivative \cite{Jain,Gonzalez,7964706}.
Therefore, $x(n)*\bar{c}_{q,k}(n)$ is equivalent to computing the first order derivative of $x(n)$.

\begin{figure}[!h]
\centering
 \includegraphics[width=4.3in,height=1.8in]{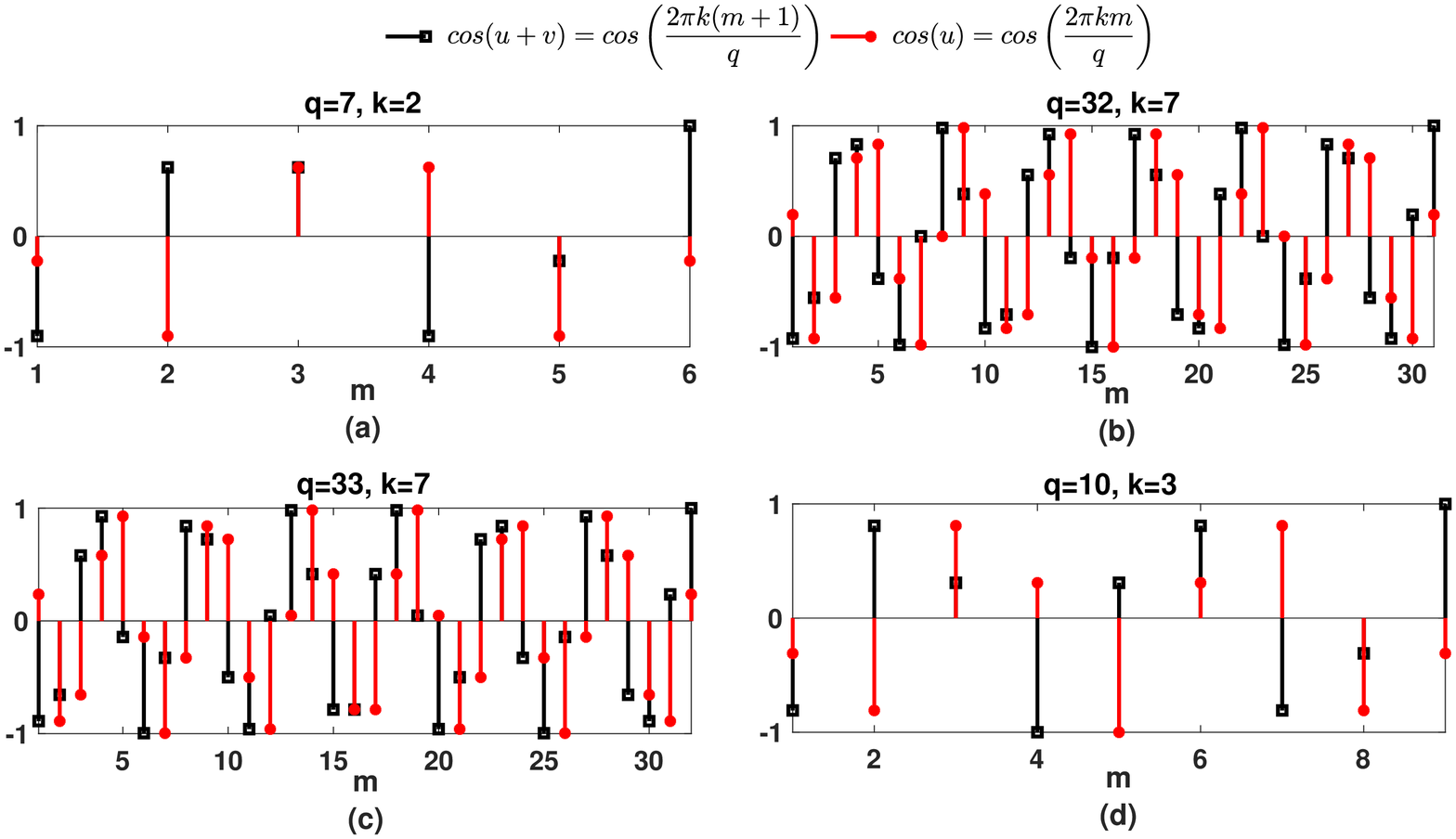} 
\caption[\small{(a)-(d) $cos\left(u+v\right)$ vs $cos\left(u\right)$ for different $q$ and $k$ values}]{\footnotesize{(a)-(d) $cos\left(u+v\right)$ vs $cos\left(u\right)$ for different $q$ and $k$ values.}}
\label{f1}
\end{figure}

Further, with some modifications in the impulse response, the above operation is  equivalent to the second order derivative.
To be a second order derivative, 
the system output should satisfy the first two conclusions mentioned in  \textbf{Theorem} \ref{Th1} and 
it should be zero for $x(n)=n$ \cite{7964706}.
That is, the term $\mathbf{P}$ in \eqref{Der1} should be equal to zero, but the assumption of $\bar{c}_{q,k}(n)$ as impulse response leads to $\mathbf{P}=Mq$. So, instead of $\bar{c}_{q,k}(n)$, we try by considering its circular shifts as an impulse response. Therefore,
\begin{equation}
\footnotesize
\sum\limits_{l=0}^{q-1}l\bar{c}_{q,k}(l-m) = \frac{q}{1-cos\left(v\right)}\left[cos\left(u+v\right)-cos\left(u\right)\right],\ 1{\leq}m{\leq}q-1,
\end{equation}
\normalsize
where $u = \frac{2{\pi}km}{q}$ and $v = \frac{2{\pi}k}{q}$. Now for what value of $m$, \footnotesize$ cos\left(u+v\right)=cos\left(u\right)$\normalsize? Figure \ref{f1} depicts \footnotesize$cos\left(u+v\right)$ vs $cos\left(u\right)\ $\normalsize  for different $q$ and $k$ values. It is observed that independent of $k$ both the values are equal whenever $q$ is an odd number and $m = \frac{q-1}{2}$.
Hence, we can summarize the above discussion as:
\begin{theorem}
For a given odd number $q$, the operation of linear convolution between a given signal $x(n)$ and $\bar{c}_{q,k}\left(n-\frac{q-1}{2}\right)$ is equivalent to computing the second order derivative of $x(n)$.
\end{theorem}
\subsection{Application}
\begin{figure}[!h]
\centerline{
 \includegraphics[width=3.3in,height=1.8in]{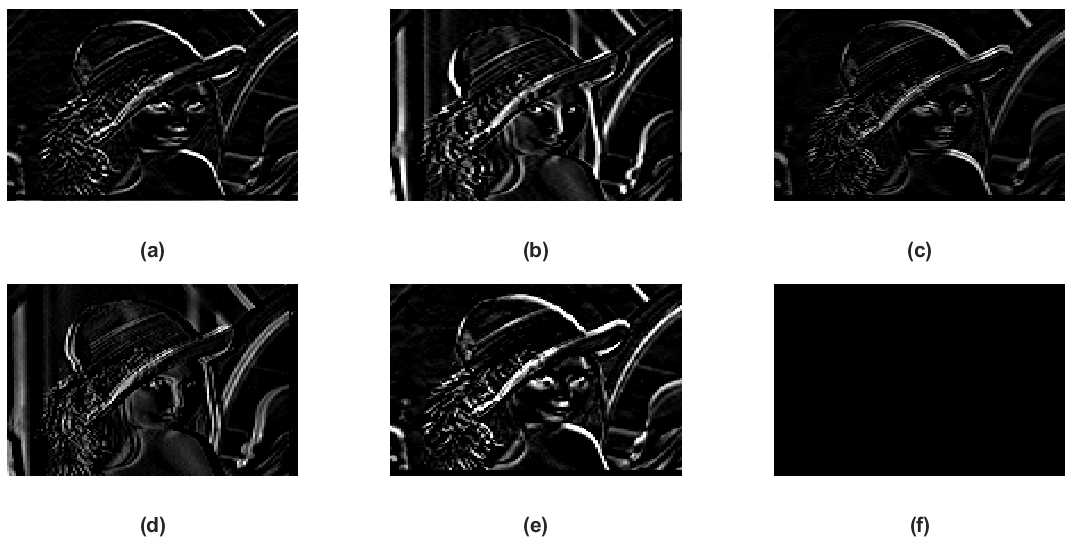} }
\caption[\small{(a)-(d) Applying convolution between Lena image and $\bar{c}_{q,k}(n)$ in both vertical (column-wise appending) and horizontal (row-wise appending) directions respectively: (a)-(b) With $q=5$ and $k=1$. (c)-(d) With $q=5$ and $k=2$. (e) Adding (a) and (c). (f) Convolving Lena image with $\bar{c}_{5}(n)$ in a vertical direction.}]{\footnotesize{(a)-(d) Applying convolution between Lena image and $\bar{c}_{q,k}(n)$ in both vertical (column-wise appending) and horizontal (row-wise appending) directions respectively: (a)-(b) With $q=5$ and $k=1$. (c)-(d) With $q=5$ and $k=2$. (e) Convolving Lena image with $\bar{c}_{5}(n)$ in a vertical direction. (f) Adding (a) and (c) then subtract the result from (e).}}
\label{f2}
\end{figure}

Edge detection of an image is crucial in many applications, where the derivative functions are used \cite{Jain,Gonzalez}. 
In this work, we address this problem using CCPSs and compare the results with the results obtained using RSs \cite{7964706}.
Consider the Lena image and convert it into two one-dimensional signals, namely $x_1(n)$ and $x_2(n)$, by column-wise appending and row-wise appending respectively.
Now compute $x_1(n)*\bar{c}_{5,1}(n),\ x_2(n)*\bar{c}_{5,1}(n),\ x_1(n)*\bar{c}_{5,2}(n)$ and $x_2(n)*\bar{c}_{5,2}(n)$, the results are depicted in figure \ref{f2} (a)-(d) respectively.
From the results, it is clear that we can find the edges using CCPSs.
Note that, performing convolution on $x_1(n)$ and $x_2(n)$ gives better detection of horizontal (Figure \ref{f2} (a) and (c)) and vertical (Figure \ref{f2} (b) and (d)) edges respectively.
Let $\hat{U}_q = \{k_1,k_2,\dots,k_{\frac{\varphi(q)}{2}}\}$, $q{\in}\mathbb{N}$, then we can write the relationship between RSs and CCPSs as
\begin{equation}
\bar{c}_q(n) = \bar{c}_{q,k_1}(n)+\bar{c}_{q,k_2}(n)+\dots+\bar{c}_{q,k_{\frac{\varphi(q)}{2}}}(n).
\end{equation}
Using this and linearity property of convolution sum, we can write $x_1(n)*\bar{c}_5(n) = x_1(n)*\bar{c}_{5,1}(n)+x_1(n)*\bar{c}_{5,2}(n)$. Figure \ref{f2} (e)-(f) validates the same.
So, fine edge detection can be achieved using CCPSs over RSs, where RSs are integer-valued sequences and CCPSs are real-valued sequences.
A similar kind of analysis can be done using CCPSs as the second derivative.

\section{Properties of CCS}
Construct a circulant matrix $\mathbf{D_{q,k}}{\in}M_{q}(\mathbb{R})$ as given below
\begin{equation}
\mathbf{D_{q,k}} = \begin{bmatrix}
\mathbf{\bar{c}}^0_\mathbf{{q,k}} & \dots & \mathbf{\bar{c}}^l_\mathbf{{q,k}}\dots & \mathbf{\bar{c}}^{q-1}_\mathbf{{q,k}}
\end{bmatrix},
\end{equation}
where $\mathbf{\bar{c}}^l_\mathbf{{q,k}}$ indicates $l$ times circular downshift of the sequence $\mathbf{\bar{c}_{q,k}}$.
Using the \textit{factorization} property of $\mathbf{D_{q,k}}$, we can write $\mathbf{D_{q,k}} = \mathbf{B_{q,k}}\mathbf{B_{q,k}^H}$, where
\begin{equation}
\mathbf{B_{q,k}^H} = \begin{bmatrix}
S_{q,{q}-k}(0)&S_{q,{q}-k}(1)&\dots&S_{q,{q}-k}(q-1) \\
S_{q,k}(0)&S_{q,k}(1)&\dots&S_{q,k}({q}-1)\\
\end{bmatrix}_{2{\times}q}.
\label{Factorization}
\end{equation}
From \eqref{Factorization}, the column space of $\mathbf{D_{q,k}}$ is equal to $v_{q,k}$. Moreover, the first two columns of $\mathbf{D_{q,k}}$ are linearly independent \cite{Shah}.
So, any signal $\mathbf{x}{\in}v_{q,k}$ can be written as,
\begin{equation}
\left[\mathbf{x}\right]_{q{\times}1} =\begin{bmatrix}
\mathbf{\bar{c}}^0_\mathbf{{q,k}}& \mathbf{\bar{c}}^1_\mathbf{{q,k}}
\end{bmatrix}\left[\mathbf{\hat{\beta}_{q,k}}\right] = \left[\mathbf{F_{q,k}}\right]_{q{\times}2}\left[\mathbf{\hat{\beta}_{q,k}}\right]_{2{\times}1}.
\label{ccsBasis}
\end{equation}
Let $q_1{\in}\mathbb{N}$, $q_2{\in}\mathbb{N}$ and $N = lcm(q_1,q_2)$, then the orthogonality property of CCPSs is \footnotesize$\left[\mathbf{\hat{c}_{q_1,k_1}}^{l_1}\right]_{N{\times}1}^T\left[\mathbf{\hat{c}_{q_2,k_2}}^{l_2}\right]_{N{\times}1}= 2NMcos\left(\frac{2{\pi}{k_1}(l_1-l_2)}{q_1}\right)\delta(q_1-q_2)\delta(k_1-k_2)$ \normalsize\cite{Shah},
where $\mathbf{\hat{c}_{q_i,k_i}}^{l_i}$ is obtained by repeating $\mathbf{\bar{c}}^{l_i}_\mathbf{{q_i,k_i}}$ periodically $\frac{N}{q_i}$ times.
Using this we can conclude the following: The vectors in the basis of $v_{q,k}$ are not orthogonal; The subspaces $v_{q_1,k_1}$ and $v_{q_2,k_2}$ are orthogonal to each other.
With this basic introduction, now we discuss a few of CCS properties in the following subsections.

%
\subsection{shift-invariant Subspace}
A natural question for a signal belongs to CCS is:
What is the effect if we consider a shifted version of the input signal?
We answer this question by proving the following theorem:
\begin{theorem}
CCS is a circular shift-invariant subspace, i.e., if $\mathbf{x}{\in}v_{q,k}$, then $\mathbf{x}^l{\in}v_{q,k}$, here the shift is interpreted as circular shift $(\text{modulo }q)$.
\end{theorem}
\textit{Proof:} From \eqref{ccsBasis}, circular shifting $\mathbf{x}$ by an amount $l$ gives 
\begin{equation}
\mathbf{x}^l = \begin{bmatrix}
\mathbf{\bar{c}}^{l}_\mathbf{{q,k}} & \mathbf{\bar{c}}^{l+1}_\mathbf{{q,k}}
\end{bmatrix}\left[\mathbf{\hat{\beta}_{q,k}}\right].
\end{equation}
$\mathbf{\bar{c}}^{l}_\mathbf{{q,k}}$ for any $l{\in}\mathbb{Z}$ is still a column in $\mathbf{D_{q,k}}$. 
It implies both $\mathbf{\bar{c}}^{l}_\mathbf{{q,k}}{\in}v_{q,k}$ and $\mathbf{\bar{c}}^{l+1}_\mathbf{{q,k}}{\in}v_{q,k}$.
This results in $\mathbf{x}^l{\in}v_{q,k}$.
In fact, using the row-Vandermonde structure of $\mathbf{B_{q,k}^H}$ one can prove that any two consecutive columns of $\mathbf{D_{q,k}}$ act as the basis for $v_{q,k}$,
hence $\mathbf{x}^l{\in}v_{q,k}$.

Since CCS is Shift-invariant, it may be useful in applications like wireless communication \cite{Forsythe},  subspace tracking:
which play a vital role in video surveillance, source localization in radar and sonar, etc., \cite{8683025}.
 
\subsection{Computing the Projections}
In CCPT, a signal $\mathbf{x}{\in}M_{N,1}(\mathbb{C})$ is represented as a linear combination of signals belongs to CCSs \cite{Shah},
\begin{equation}
\mathbf{x} = \sum\limits_{q_i|N}^{}\sum\limits_{\substack{k=1\\(k,q_i)=1}}^{\floor*{\frac{q_i}{2}}}\underbrace{\mathbf{E_{q_i,k}}\boldsymbol{\beta}_\mathbf{{q_i,k}}}_{\mathbf{x_{q_i,k}}{\in}v_{q_i,k}} = \left[\mathbf{T_N}\right]_{N{\times}N}\left[\boldsymbol{\beta}\right]_{N{\times}1}.
\label{Synthesis}
\end{equation}
Here, $\mathbf{T_N}$ is the transformation matrix, $\boldsymbol{\beta}$ is the transform coefficient vector,
$\mathbf{x_{q_i,k}}$ denotes the projection of $\mathbf{x}$ onto $v_{q_i,k}$,
$\mathbf{[E_{q_i,k}]_{N{\times}2}}= [\mathbf{F_{q_i,k}},\dots,\mathbf{F_{q_i,k}}]^T$ 
and $\boldsymbol{\beta}_\mathbf{{q_i,k}}{\in}M_{2,1}(\mathbb{C})$ is the transform coefficients vector corresponds to $v_{q_i,k}$. 
As given in \cite{Shah}, a major application of CCPT is 
estimating the period and frequency information of a signal using $\boldsymbol{\beta}$.
While the non-orthogonal basis of $v_{q_i,k}$ makes CCPT a non-orthogonal transform, i.e., it requires computation of $\mathbf{T_N^{-1}}$ to find $\boldsymbol{\beta}$. 
According to Strassen's algorithm, computing $\mathbf{T}^{-1}_\mathbf{N}$ requires a computational complexity of $\mathcal{O}(N^{2.81})$ \cite{Cormen}.
To use CCPT in an efficient way, one has to overcome this limitation, for this
we compute $\mathbf{x_{q_i,k}}$ instead of $\boldsymbol{\beta}$, which can serve for the same purpose.
Since $\left[\mathbf{E_{q_i,k}}\right]^H\left[\mathbf{E_{q_x,k_y}}\right] = 0,\ \forall\ i{\neq}x$ (where $q_i|N \text{ and } q_x|N$), the projection of $\mathbf{x}$ onto $v_{q_i,k}$ can be computed as follows:
\begin{equation}
\mathbf{x_{q_i,k}} = \underbrace{\mathbf{E_{q_i,k}}(\mathbf{E_{q_i,k}^H}\mathbf{E_{q_i,k}})^{-1}\mathbf{E_{q_i,k}^H}}_{\mathbf{P_{q_i,k}}{\in}M_{N}:\text{ Projection matrix}}\mathbf{x}.
\label{projMatrix}
\end{equation}
Here $\mathbf{P_{q_i,k}}$ can be further reduced as given below:
\begin{equation}
\mathbf{P_{q_i,k}} = \frac{q_i}{N}\begin{bmatrix}
\mathbf{\hat{P}_{q_i,k}}&\dots&\mathbf{\hat{P}_{q_i,k}}\\
\vdots&\ddots&\vdots\\
\mathbf{\hat{P}_{q_i,k}}&\dots&\mathbf{\hat{P}_{q_i,k}}
\end{bmatrix}, \text{where}
\label{redProj}
\end{equation} 
\begin{equation}
\mathbf{[\hat{P}_{q_i,k}]_{q_i{\times}q_i}} = \mathbf{F_{q_i,k}}(\mathbf{F_{q_i,k}^H}\mathbf{F_{q_i,k}})^{-1}\mathbf{F_{q_i,k}^H}.
\end{equation}
Let us divide the input signal $\mathbf{x}$ into $\frac{N}{q_i}$ blocks, that is $\mathbf{x} = \left[\mathbf{x^{(1)}},\mathbf{x^{(2)}},\dots,\mathbf{x^{(\frac{N}{q_i})}}\right]^T$, where $i^{th}$ block $\mathbf{x^{(i)}}{\in}M_{{q_i}{\times}1}$.
Then, from \eqref{projMatrix} and \eqref{redProj} we can write
\begin{equation}
\mathbf{x_{q_i,k}} = \frac{q_i}{N}\begin{bmatrix}
\mathbf{\hat{P}_{q_i,k}}&\dots&\mathbf{\hat{P}_{q_i,k}}\\
\vdots&\ddots&\vdots\\
\mathbf{\hat{P}_{q_i,k}}&\dots&\mathbf{\hat{P}_{q_i,k}}
\end{bmatrix}\begin{bmatrix}
\mathbf{x^{(1)}}\\
\vdots\\
\mathbf{x^{(\frac{N}{q_i})}}
\end{bmatrix} = \begin{bmatrix}
\mathbf{y_{q_i,k}}\\
\vdots\\
\mathbf{y_{q_i,k}}
\end{bmatrix},
\label{Proj}
\end{equation}
where
\begin{equation}
\mathbf{y_{q_i,k}} =\frac{q_i}{N}\mathbf{\hat{P}_{q_i,k}}\sum\limits_{i=1}^{\frac{N}{q_i}}\mathbf{x^{(i)}}.
\label{Proj1}
\end{equation}
Now consider a matrix $\mathbf{\tilde{P}_{q_i,k}} = \frac{\mathbf{D_{q_i,k}}}{q_i}$, then $\mathbf{\tilde{P}_{q_i,k}}^2 = \frac{\mathbf{B_{q_i,k}}\mathbf{B}^H_\mathbf{{q_i,k}}\mathbf{B_{q_i,k}}\mathbf{B}^H_\mathbf{{q_i,k}}}{q_i^2} = \mathbf{\tilde{P}_{q_i,k}}$, since $\mathbf{B}^H_\mathbf{{q_i,k}}\mathbf{B_{q_i,k}} = q_i\mathbf{I}$.
Moreover, the even symmetry property of CCPSs makes $\mathbf{\tilde{P}_{q_i,k}}^H = \mathbf{\tilde{P}_{q_i,k}}$.
So, we can say that $\mathbf{\tilde{P}_{q_i,k}}$ is an orthogonal projection matrix. 
Since $v_{q_i,k}$ is the column space of $\mathbf{D_{q_i,k}}$, 
we can write $\mathbf{\hat{P}_{q_i,k}} = \mathbf{\tilde{P}_{q_i,k}}= \frac{\mathbf{D_{q_i,k}}}{q_i}$.
With this, \eqref{Proj1} can be modified as,
\begin{equation}
\mathbf{y_{q_i,k}} =\frac{1}{N}\mathbf{D_{q_i,k}}\sum\limits_{i=1}^{\frac{N}{q_i}}\mathbf{x^{(i)}} =\frac{1}{N}\mathbf{D_{q_i,k}}\mathbf{\hat{x}} .
\label{Proj2}
\end{equation}
So, the orthogonal projection $\mathbf{x_{q_i,k}}$ is computed as follows: First, multiply $\mathbf{\hat{x}}$ with the circulant matrix $\mathbf{D_{q_i,k}}$, then multiply the result with a scale factor $\frac{1}{N}$ to get $\mathbf{y_{q_i,k}}$. Now repeat $\mathbf{y_{q_i,k}}$ periodically $\frac{N}{q_i}$ times to obtain $\mathbf{x_{q_i,k}}$.
The number of multiplications (computational complexity) involved in computing $\mathbf{x_{q_i,k}}$ are ${q_i}^2+q_i$.
From \eqref{Synthesis}, there are total of $\frac{1}{2}\sum\limits_{q_i|N}^{}\varphi(q_i) = \frac{N}{2}$ number of CCSs are involved in representing $\mathbf{x}$. 
So, the total number of multiplications ($M_{total}$) required for computing projections for all $\frac{N}{2}$ CCSs are 
\begin{equation}
M_{total} = \begin{cases}
2+\frac{1}{2}\sum\limits_{\substack{q_i|N\\ q_i{\geq}3}}^{}\varphi(q_i)(q_i^2+q_i),\ &\text{if }2\nmid N \\
8+\frac{1}{2}\sum\limits_{\substack{q_i|N\\ q_i{\geq}3}}^{}\varphi(q_i)(q_i^2+q_i),\ &\text{if }2\mid N 
\end{cases}.
\end{equation}
\begin{table}[h]
\centering
\caption{C\scriptsize{OMPARISON OF $M_{total}\ vs\ \floor*{N^{2.81}}$ FOR FEW $N$ VALUES}}
\label{tab:Comparison}
\begin{adjustbox}{max width=\textwidth}
\scalebox{0.8}{
\begin{tabular}{|M{1.1cm}|M{1cm}|M{1.2cm}|M{1.4cm}|M{1.4cm}|M{1.4cm}|} \hline
N &	$3$	&	$6$ & $8$&	$32$	& $82$	\\	\hline
\small{$M_{total}$} &	$14$	&	$62$ &	$172$& $9708$	&  $170568$  \\	\hline
\small{$\floor*{N^{2.81}}$} & $21$	&	$153$ &$344$&	$16961$	& $238680$  \\    \hline
\end{tabular}}
\end{adjustbox}
\end{table}
For few $N$ values both $M_{total}$ and $\floor*{N^{2.81}}$ are tabulated in Table \ref{tab:Comparison}. 
From the table, we can conclude that $M_{total} << \mathcal{O}(N^{2.81})$ and there is an approximate of $40\%$ reduction in the computational complexity for the values given in the table.
Therefore, estimating the period and frequency information through
projection computation requires less computational complexity over the direct inverse computation method.
Apart from this computational advantage, in some applications, it is required to know the existence of certain periods in an observed signal.
In such scenarios, computing projections for those CCSs are sufficient.
\subsection{Correlation of Sequences in CCS}
\textbf{DFT of CCPS:} For a given $q{\in}\mathbb{N}$ and $k{\in}\hat{U}_q$,
\begin{equation}
DFT[\bar{c}_{q,k}(n)] = \bar{C}_{q,k}(K) = \begin{cases}
q,\ \text{if }K=k\ \text{(or)}\ q-k,\\
0,\ \text{Otherwise}.
\end{cases}
\end{equation}
Let $r_{c}(l)$ denotes the circular autocorrelation of $\bar{c}_{q,k}(l)$, then
\begin{equation}
\nonumber
DFT[r_{c}(l)] = R_{c}(K) = \bar{C}_{q,k}(K)\bar{C}_{q,k}(K) = q\bar{C}_{q,k}(K).
\end{equation}
Taking IDFT of above equation leads to $r_{c}(l) = q\bar{c}_{q,k}(l)$.
Using this relation, we prove the following Theorem.
\begin{theorem} CCS is closed under circular cross-correlation operation, i.e., if $x(n){\in}v_{q,k}$ and $y(n){\in}v_{q,k}$ then the circular cross-correlation $r_{xy}(l)$ also belongs to $v_{q,k}$.
\end{theorem}
\textit{Proof:} Given $x(n){\in}v_{q,k}$ and $y(n){\in}v_{q,k}$ then
\begin{equation}
\nonumber
\footnotesize
\begin{aligned}
r_{xy}(l)& = \sum\limits_{n=0}^{q-1}\underbrace{\left[\sum\limits_{m_1=0}^{1}{\beta_{m_1}}\bar{c}_{q,k}(n-m_1)\right]}_{x(n)}\underbrace{\left[\sum\limits_{m_2=0}^{1}{\gamma^*_{m_2}}\bar{c}_{q,k}(n-l-m_2)\right]}_{y^*(n-l)}\\
& = \sum\limits_{m_1=0}^{1}\sum\limits_{m_2=0}^{1}{\beta_{m_1}}{\gamma^*_{m_2}}\left[\sum\limits_{n=0}^{q-1}\bar{c}_{q,k}(n-m_1)\bar{c}_{q,k}(n-l-m_2)\right]\\
& = \sum\limits_{m_1=0}^{1}\sum\limits_{m_2=0}^{1}{\beta_{m_1}}{\gamma^*_{m_2}}q\underbrace{\bar{c}_{q,k}(l+m_2-m_1)}_{{\in}v_{q,k}}.
\end{aligned}
\normalsize
\end{equation}
From above, $r_{xy}(l)$ is a weighted linear combination of signals belongs to $v_{q,k}$, hence $r_{xy}(l){\in}v_{q,k}$.

Further, the autocorrelation of a $N$-length signal $x(n)$ is
\begin{equation}
r_x(l) = \sum\limits_{n=0}^{N-1}x(n)x^*(n-l)
\end{equation}
Using \eqref{Synthesis}, the above equation can be modified as
\footnotesize
\begin{equation}
r_x(l) = \sum\limits_{q_i|N}^{}\sum\limits_{\substack{k_1=1\\(k_1,q_i)=1}}^{\floor*{\frac{q_i}{2}}}\sum\limits_{q_j|N}^{}\sum\limits_{\substack{k_2=1\\(k_2,q_j)=1}}^{\floor*{\frac{q_j}{2}}}\underbrace{\left[\sum\limits_{n=0}^{N-1}x_{q_i,k_1}(n)x^*_{q_j,k_2}(n-l)\right]}_{\mathbf{Q}}.
\label{Corr}
\end{equation}
\normalsize
where $\mathbf{Q} = \begin{cases}
\frac{N}{q_i}r_{x_{q_i,k_1}}(l),&\ \text{if } q_i=q_j\ \&\ k_1=k_2\\
0,&\ \text{if }{k_1}{\neq}{k_2}
\end{cases},
$
since CCS is shift-invariant and both $v_{q_i,k_1}$, $v_{q_j,k_2}$ are orthogonal to each other  for $k_1{\neq}k_2$.
Now substituting $\mathbf{Q}$ in \eqref{Corr} leads to
\begin{equation}
\footnotesize
\frac{1}{N}r_x(l) = \sum\limits_{q_i|N}^{}\sum\limits_{\substack{k_1=1\\(k_1,q_i)=1}}^{\floor*{\frac{q_i}{2}}}\frac{1}{q_i}r_{x_{q_i,k_1}}(l).
\normalsize
\end{equation}
The above result is summarized as follows:
\begin{theorem}
The circular autocorrelation of a signal is
equal to the linear combination of 
its  projections (onto CCSs) autocorrelation, with a proper normalization.
\end{theorem}
This property is useful if the given signal is contaminated with additive noise. In such scenarios projecting $r_x(l)$ onto CCSs gives an accurate period and frequency estimation over projecting $x(n)$.

\section{Conclusion}
In this work, we have shown how to use CCPSs as first and second order derivatives. Here, the first order derivative is used to find the edges in an image.
It is shown that fine edge detection can be achieved using CCPSs over RSs.
Later we discussed few properties of CCSs, which may find their applications in signal processing.
In particular, the signal information can be estimated with lesser computational complexity using CCS projections.
\section*{Acknowledgement}
The authors would like to thank Mr. Shiv Nadar, the founder and chairman of HCL and Shiv Nadar Foundation.

\ifCLASSOPTIONcaptionsoff
  \newpage
\fi



\bibliographystyle{ieeetr}
\bibliography{bibs_1}
\end{document}